\documentclass[twocolumn]{revtex4}
\usepackage{graphicx,graphics}
\usepackage{amsmath,epsf,subfigure}
\usepackage{color}
\usepackage{amsfonts,amsmath,graphics}
\usepackage{bm,subfigure}
\usepackage{graphicx,epsf}
\usepackage{amssymb}
\usepackage{afterpage}
\usepackage{float}
\usepackage{chemmacros}
\usepackage{tikz-cd}
\usepackage{MnSymbol}
\usepackage{amsmath}
\usepackage{comment}

\renewcommand{\:}{{:}}
\newcommand{\dd}{{\ddagger}}
\renewcommand{\S}{\mathcal{S}}

\newcommand{\beq}{\begin{equation}}
\newcommand{\eeq}{\end{equation}}
\newcommand{\harp}{\xrightleftharpoons}
\newcommand{\xa}{\xrightarrow[]}

\renewcommand{\b}{\beta}

\renewcommand{\l}{\lambda}
\newcommand{\e}{\epsilon}

\renewcommand{\r}{\rho}

\begin{document}

\title{One-way catalysis in a solvable lattice model}

\author{Sara Mahdavi}
\author{Yann Sakref}
\author{Olivier Rivoire}

\affiliation{Gulliver, ESPCI, CNRS, Universit\'e PSL, Paris, France.}

\begin{abstract}

Catalysts speed up chemical reactions with no energy input and without being transformed in the process, therefore leaving equilibrium constants unchanged. Some catalysts, however, are much more efficient at accelerating one direction of a reaction. Is it possible for catalysis to be strictly unidirectional, accelerating only one direction of a reaction? Can we observe directional catalysis by analyzing the microscopic trajectory of a single reactant undergoing conversions between a substrate and a product state? We use the framework of a simple but exactly solvable lattice model to study these questions. The model provides examples of strictly one-way catalysts and illustrates a mathematical relationship between the asymmetric transition rates that underlie directional catalysis and the symmetric transition fluxes that underlie chemical equilibrium. The degree of directionality generally depends on the catalytic mechanism and we compare different mechanisms to show how they can obey different scaling laws. 
\end{abstract}

\maketitle

\section{Introduction}

Catalysis, the acceleration of chemical reactions by substances that remain unchanged, is central to biological processes as well as the chemical industry. It also plays a key role in many non-equilibrium phenomena studied in statistical physics, from molecular motors to active matter. In these contexts, catalysis is most often treated phenomenologically by imposing kinetic rates without considering the microscopic interactions that underlie them. This allows one to ignore the chemical mechanisms by which catalysis operates, but at the cost of leaving aside any question pertaining to the physical limitations of these mechanisms.

To address such questions, we have previously introduced simplified models of catalysis in which the kinetic states and rates of a catalytic cycle are derived from a microscopic description of interacting particles~\cite{Rivoire.2020,Munoz-Basagoiti.2023,Rivoire.2023}. A first class of models, in which the catalyst consists of rigidly held particles, reproduces two general principles of chemical catalysis~\cite{Rivoire.2020,Munoz-Basagoiti.2023}: Pauling's principle stating that the geometry of an optimal catalyst should be complementary to the transition state of the reaction~\cite{PAULING.1948}, and Sabatier's principle stating that the interaction between the reactants and the catalyst should be neither too weak nor too strong~\cite{gor}. For this class of models, we found that catalysis can only lower the activation barrier of a reaction by a finite factor~\cite{Rivoire.2023}, consistent with observations in chemistry~\cite{Rivoire.2024}. A second class of models, in which the catalyst can switch between two internal states, reproduces the ability of enzymes to overcome this limitation by an allosteric mechanism~\cite{Rivoire.2023,Rivoire.2024}.

Here we apply and extend our approach to examine a simple question: if a substance is a catalyst for one direction of a reaction, is it necessarily a catalyst for the opposite direction of the same reaction? The principle that catalysis does not affect equilibrium properties is sometimes misinterpreted to imply that catalysis must be inherently bidirectional~\cite{vsima2015catalysis,chorkendorff2017concepts,ChemLibrText}, but directional catalysis, where catalysts preferentially accelerate one direction of a reaction, is well documented in enzymology, where it is referred to as one-way catalysis~\cite{Jencks.1993b18}, as well as in chemistry, where it is referred to as catalytic bias~\cite{fourmond2019understanding}. It has been repeatedly noted that such one-way catalysis does not contradict any fundamental principle: catalysts accelerate only reactions that are thermodynamically favorable, and the two directions of a reaction are favorable under different conditions, with either a large concentration of substrates or a large concentration of products~\cite{Jencks.1993b18,cornish2013fundamentals}. However, this argument does not reveal what happens at the microscopic level of individual molecules. In particular, at chemical equilibrium, while the relative concentration of reactants and products remains globally unchanged, the presence of catalysts must affect the frequency of microscopic transitions between substrates and products by the same factor. How does the preferential acceleration of one direction of reaction relate to this symmetric effect on transition frequencies?

To clarify this point, we draw on two elements of our previous work. First, we study catalysis in the framework of a simple lattice model that can be solved without approximation by exact numerical calculations and, in particular cases, analytical calculations~\cite{Rivoire.2023}. We analyze the simplest reaction that can be defined in this framework: the dissociation or association of two particles interacting with a short-range potential. Second, we apply a quantitative definition of catalysis based on first-passage times~\cite{Ninio.1987,Qian.20080tj,Munoz-Basagoiti.2023}, which we have shown to be consistent with other common definitions~\cite{Sakref.2023}. In this framework, reaction rates, usually obtained from macroscopic changes in substrate and product concentrations, are obtained as inverse mean first-passage times for a single reactant undergoing interconversions between a substrate and a product state. As we illustrate, this single-particle view can reveal features not apparent at the macroscopic level, including changes in transition rates that do not affect global concentrations, or fluctuations captured by the higher moments of first-passage time distributions.

Using these approaches and notions from transition path theory~\cite{Metzner.2009}, we show that chemical equilibrium and directional catalysis are respectively encoded in microscopic trajectories through reaction fluxes, which are symmetric at chemical equilibrium, and reaction rates, which are generally asymmetric. In particular, we show that it is possible to have strictly one-way catalysis, where the rate of one side of a reaction is accelerated but not the other. The demonstration is based on a reaction that has two activation barriers of different nature, energetic and entropic. This leads us to examine how different catalytic mechanisms are suitable for lowering different types of activation barriers, and how they exhibit different scaling laws as a function of the amplitude of these activation barriers.

\section{Methods}

For the sake of completeness, we first present the modeling framework on which we rely, first introduced in~\cite{Rivoire.2023}. A compendium of useful mathematical formulas is provided in the Appendices.

\subsection{Microscopic model}

\begin{figure}[t]
\begin{center}
\includegraphics[width=.9\linewidth]{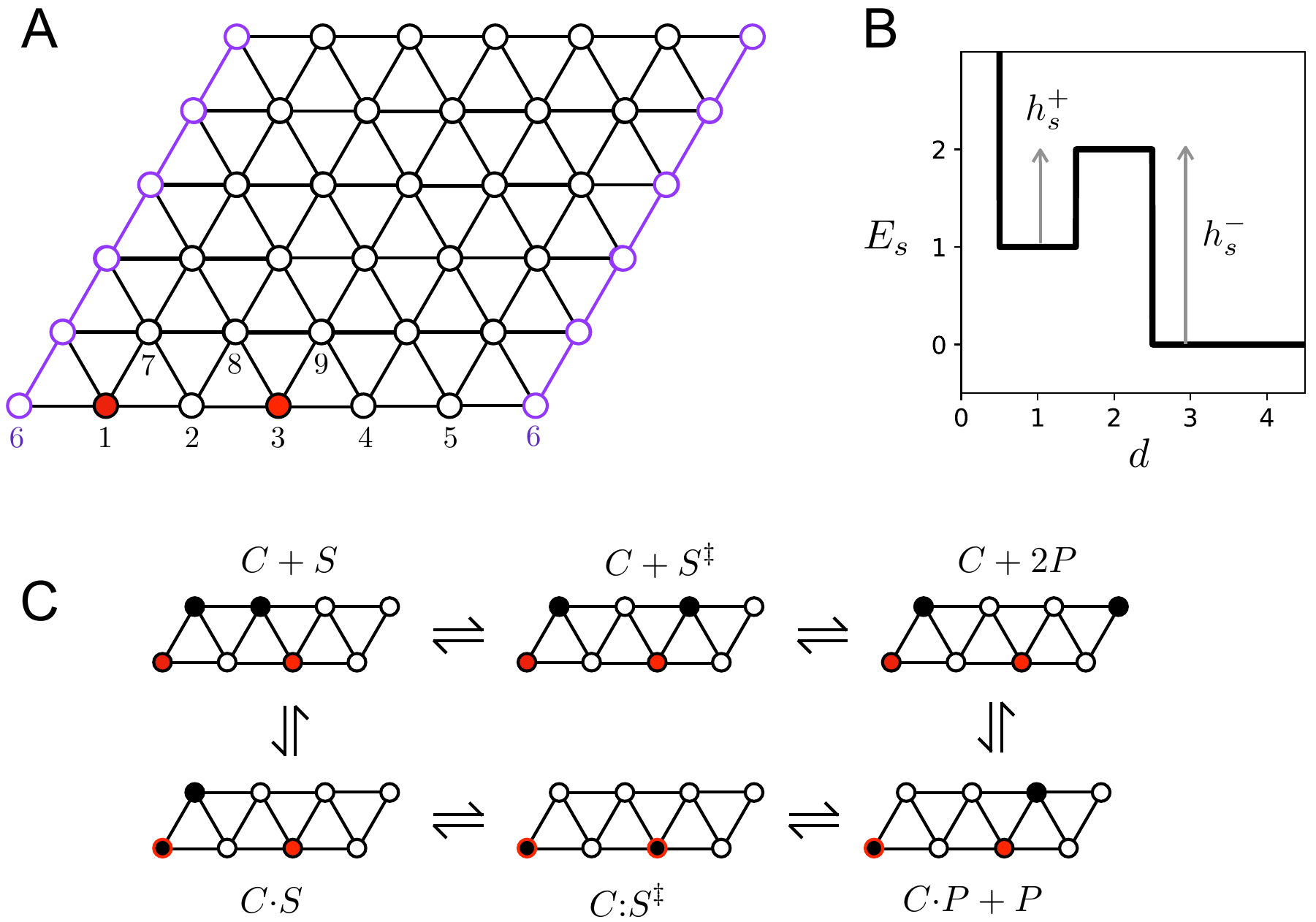}
\caption{\small Model definition. {\bf A.} A two-dimensional lattice of size $L\times L$ with periodic boundary conditions along the $x$ axis (repeated purple nodes). The lattice nodes represent spatial locations that a particle can occupy if they are not already occupied by another particle. The nodes are labeled $i=1,\dots,L^2$. Some nodes $\S=\{j_1,\dots,j_n\}$, e.g. $\S=\{1,3\}$ shown in red, are defined as binding sites with associated binding energies $\e_{cs}^{j_k}$. If a particle is at the binding site $j_k$, the energy is lowered by $\e_{cs}^{j_k}$. {\bf B.} The potential of interaction between two particles. This potential is defined by the distance $d$ between the two particles, measured by the minimal path separating them. Particles at distance $d=1$ are in a bound state $S$, and particles at distance $d\geq 3$ are in a free state $2P$. The potential is parameterized by the two activation energies $h_s^+$ and $h_s^-$. Here $h_s^+\leq h_s^-$, but the opposite is also possible. {\bf C.}~We group configurations into macrostates: for instance, configurations where the two particles (in black) neither interact with each other nor with the binding sites (in red) are represented by $C+S$ (which stands for catalyst $+$ substrate). In total, we distinguish six macrostates, illustrated here by a configuration that they contain for the case of a catalyst consisting of two binding sites at distance $d=2$. The possible transitions between these different macrostates define a catalytic cycle. The upper configurations in our depiction of this cycle, in which the particles do not occupy the binding sites, represent the spontaneous reaction. The lower configurations represent the macrostates with an interaction with the catalyst: $C\.S$ corresponds to a bound state with a particle on a binding site, $C\:S^\dd$ to a transition state with the two binding sites occupied, and $C\.P+P$ to a state with one bound particle and one free particle. \label{fig:scheme}}
\end{center} 
\end{figure}

We consider a model of particles occupying the sites of a two-dimensional hexagonal lattice with periodic boundary conditions along one direction, as shown in Fig.~\ref{fig:scheme}A. We restrict our study to two non-overlapping particles that can diffuse from one site to the next and form a dimer when they are close together. The interaction between the particles is determined by the potential shown in Fig.~\ref{fig:scheme}B, which depends only on the distance $d$ between the two particles, measured by the minimum number of sites separating them on the lattice. Particles at distance $d=1$ are in a stable bound state denoted $S$ (which stands for substrate), particles at distance $d\geq 3$ are in a non-interacting state denoted $P$ (which stands for product), while particles at distance $d=2$ are in an unstable transition state denoted $S^\ddagger$. Paths from $S$ to $P$, which must necessarily go through $S^\ddagger$, thus represent a ``reaction'' of dimer association. Each of these macrostates consists of multiple microscopic configurations in which the particles are located at different lattice nodes. The potential has two parameters: a dissociation barrier $h_s^+$ and an association barrier $h_s^-$ (Fig.~\ref{fig:scheme}B); in other words, $h_s^+$ represents the activation energy for the reaction of dissociation of a dimer into two monomers and $h_s^-$ for the reverse reaction.

A configuration $x=(i_1,i_2)$ with the two particles at positions $i_1$ and $i_2$ thus has an interaction energy given by
\beq
E_s(i_1,i_2) = \left\{
    \begin{array}{ll}
        +\infty & \mbox{if } d(i_1,i_2)=0 \\
       h_s^--h_s^+ & \mbox{if } d(i_1,i_2)=1\\
       h_s^- & \mbox{if } d(i_1,i_2)=2\\
       0 & \mbox{if } d(i_1,i_2)\geq 3\\
    \end{array}
\right.
\eeq

A rigid and fixed catalyst is defined by assigning to some sites $\S=\{j_1,\dots,j_n\}$ of the lattice binding energies $\e_{cs}^{j_1},\dots, \e_{cs}^{j_2}$ such that when a particle occupies site $j_k$, its energy is lowered by $\e_{cs}^{j_k}$. As shown in previous works~\cite{Rivoire.2020,Munoz-Basagoiti.2023,Rivoire.2023}, the simplest configuration providing catalysis consists of two binding sites at distance $d=2$ with a common binding energy $\e_{cs}$, for example $\S=\{1,3\}$ (red sites in Fig.~\ref{fig:scheme}A) with $\e_{cs}^{1}=\e_{cs}^{3}=\e_{cs}$. In general, the total energy of the configuration $x=(i_1,i_2)$ is 
\beq
E(i_1,i_2)=E_s(i_1,i_2)-\sum_{j\in\S}(\e_{cs}^j\delta_{j,i_1}+\e_{cs}^j\delta_{j,i_2})
\eeq
where $\delta_{j,i_k}=1$ if $i_k=j$ and $0$ otherwise.

To study catalysis of the dissociation reaction, we start with two particles are in a bound ``substrate'' state $S$ (defined by a distance $d=2$ between the 2 particles), away from the binding sites. They can then diffuse across the lattice nodes and may come to occupy one or two of the particular nodes of the lattice that define the binding sites of the catalyst (red nodes in Fig.~\ref{fig:scheme}A). The dissociation of $S$ into a product state $P$ (defined by a distance $d\geq 3$ and no interaction with the binding sites) can occur in two ways. It can occur spontaneously, without any interaction with the binding sites, through a high-energy transition state $S^\ddagger$ (where $d=2$). Alternatively, it can occur through interaction with the binding sites. In a particular case where the catalyst consists of two binding sites at a distance of $d=2$, the catalytic process first involves a state where only one molecule occupies one binding site (state $C\.S$) before the two molecules occupy the two binding sites (state $C\:S^\dagger$). The activation energy of $C\.S\to C\:S^\ddagger$ is then reduced by $\e_{cs}$ compared to $S\to S^\ddagger$. From $C\:S^\ddagger$, one molecule can then diffuse out of a binding site (state $CP+P$), followed by the other (state $C+2P$). However, all transitions are reversible and catalysis may involve multiple binding and unbinding events before the full reaction $C+S\to C+2P$ is completed. In the following, we also study the reverse scenario, starting from $C+2P$ and ending in $C+S$, as well as different configurations of the binding sites.

More formally, transitions can occur as a result of thermal fluctuations from one configuration $x$ of the particles to another $y$ where one of the particles has moved to a neighboring site. Each configuration consists of the positions on the lattice of the two particles, indicated by the labels $(i_1,i_2)$ of the nodes that they occupy. If $x=(i_1,i_2)$, we can have either $y=(i'_1,i_2)$ with $d(i'_1,i_1)=1$ and $d(i'_1,i_2)>0$, or $y=(i_1,i'_2)$ with $d(i'_2,i_2)=1$ and $d(i'_2,i_1)>0$. We assume that these microscopic transitions occur at a rate given by
\beq\label{eq:rho}
\r(x\to y)=\rho_0\min(1,e^{-\b (E(y)-E(x))})
\eeq
where $\rho_0$ defines a diffusion rate, which we set to $1$ in the following, and where $\beta=1/(k_B T)$ is the inverse of the temperature multiplied by the Boltzmann constant. This choice of transition rates, corresponding to the Metropolis rule, ensures that the dynamics satisfies detailed balance, thereby guaranteeing that the system asymptotically reaches an equilibrium distribution governed by the Boltzmann law, where the probability $\pi(x)$ of being in the configuration $x$ is proportional to $e^{-\b E(x)}$. 

To sum up, the parameters of the microscopic model include the geometry of the lattice, which we take to be of dimension $L\times L$ ($L=6$ in Fig.~\ref{fig:scheme}A), the two activation energies $h_s^+$ and $h_s^-$ for the spontaneous reaction, the choice of the number and location of the binding sites $\S$ and their energies $\e_{cs}^j$ which define a putative catalyst, and the inverse temperature $\b$. While the latter could be fixed to 1 without loss of generality, it is convenient to keep it as an explicit parameter to study scaling laws in the limit of a low temperatures ($\b\to\infty$). Unless otherwise mentioned, we take $h_s^+=1$, $h_s^-=2$, $\e_{cs}=1/2$ and $\b=10$ for the figures.

\subsection{Definition and quantification of catalysis}

Following~\cite{Munoz-Basagoiti.2023,Sakref.2023,Rivoire.2023}, we define and quantify catalysis by the mean first-passage time to complete a reaction. This is motivated by the identification of reaction rates, commonly studied in chemical kinetics, with inverse mean first-passage times~\cite{Ninio.1987}. We focus first on the cleavage reaction by which a dimer of two particles at distance $d=1$ dissociates into two free particles at distance $d\geq 3$. To formally define this reaction, we collect all configurations in which the particles form a dimer that does not interact with the catalyst, i.e, configurations $x=(i_1,i_2)$ with $d(i_1,i_2)=1$ and $i_1,i_2\notin \S$ into a macrostate $C+S$, and all configurations where the two particles are free and not interacting with the catalyst, i.e, configurations $x=(i_1,i_2)$ with $d(i_1,i_2)\geq 3$ and $i_1,i_2\notin \S$ into a macrostate $C+2P$. A catalyzed reaction, denoted $C+S\to C+2P$, is achieved through a path from a configuration in $C+S$ to a configuration in $C+2P$. Starting from a given configuration $x$ in $C+S$, the time to first reach a configuration in $C+2P$ based on Eq.~\eqref{eq:rho} is stochastic. The average time over stochastic trajectories defines the mean first-passage time $T_{x\to C+2P}$. Further averaging over the configurations $x$ in $C+S$ defines the global mean first-passage time $T_{C+S\to C+2P}$. This definition can be extended to define the moment of order $n$ of the first-passage time, with the mean first-passage time corresponding to the first moment $n=1$.

When all interaction energies $\e_{cs}^{j_k}$ are zero, so that the binding sites do not differ from other lattice sites, the mean first-passage time indicates the mean time to complete the spontaneous reaction and is simply denoted $T_{S\to P}$~\footnote{In previous work~\cite{Munoz-Basagoiti.2023}, we included in the states $S$ and $2P$ configurations in which one or two particles occupy a noninteracting binding site. For large volumes, the difference is negligible, but excluding the occupation of noninteracting binding sites allows a more meaningful comparison with the catalyzed reaction independent of volume.}. We say that a particular choice of interaction energies defines a catalyst if $T_{C+S\to C+2P}$ is less than $T_{S\to 2P}$. The ratio between these two times, denoted $\eta_{C+S\to C+2P}=T_{S\to 2P}/T_{C+S\to C+2P}$, which must be $>1$ for catalysis to occur, quantifies the catalytic efficiency. We have previously shown how this definition of catalysis is consistent with other definitions in chemistry and enzymology, and how, for the particular cleavage reaction $S\to 2P$, performing the analysis with two particles is sufficient to cover catalysis with more particles~\cite{Sakref.2023}.

\subsection{Calculations}

Several approaches are available to compute the moments of first-passage times and thus verify whether a particular choice of binding sites and their interaction energy defines a catalyst. One possibility is to perform numerical simulations of stochastic trajectories using a kinetic Monte Carlo algorithm~\cite{Chatterjee.2007}. This approach is not exact, but it is very general. The main difficulty is that the time for the simulations diverges with increasing values of $\b$. Alternative numerical approaches have been developed to get around this problem~\cite{Wales.2009}. Here we follow~\cite{Rivoire.2023} and compute the mean first passage directly by algebraic manipulation of the transition matrix of the associated Markov process, which amounts to directly solving the master equation or, more precisely, the backward Kolmogorov equation (see Appendices). This approach is exact to numerical precision, but is only suitable for relatively small lattice sizes, since it involves inverting a matrix whose size scales with the total number of configurations. If $L^2$ is the number of lattice sites, then the number of different configurations $x$ is indeed $N=L^2(L^2-1)$, which scales as $L^4$. Since the questions that we are investigating only require a lattice size sufficient to place four consecutive binding sites at the bottom (the maximal number of binding sites considered in this work), this size can be small and we take $L=6$ here, which allows us to make this approach tractable. In addition to these exact numerical calculations, we also analyze coarse-grained descriptions of the models in the form of Markov processes between the different macrostates, which allows for analytical results.

\begin{figure}[t]
\begin{center}
\includegraphics[width=\linewidth]{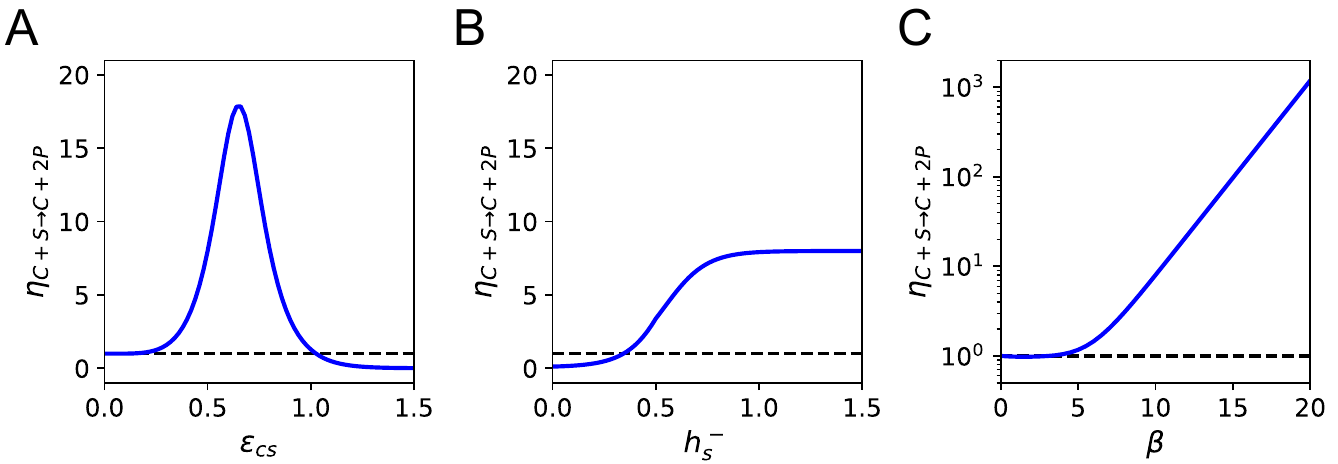}
\caption{\small Catalytic efficiency $\eta_{C+S\to C+2P}$ for a catalyst consisting of two sites at distance $d=2$, here shown for $h_s^+=1$, $h_s^-=2$, $\e_{cs}=1/2$ and $\b=10$, unless one of these parameters is varied. Catalysis occurs when $\eta_{C+S\to C+2P}>1$, a threshold indicated by the horizontal dashed black line. {\bf A.}~Dependence on the interaction $\e_{cs}$, showing a maximum at an intermediate value of $\e_{cs}$. This graph is analogous to Volcano plots drawn from empirical data in heterogeneous catalysis and reflects the Sabatier principle~\cite{Medford.2015}: too weak an interaction is insufficient to accelerate the chemical step $C\.S\to C\.P+P$, while too strong an interaction prevents efficient release $C\.P+P\to C+2P$. {\bf B.}~Dependence on the reverse activation energy $h_s^-$, showing that catalysis requires a sufficiently large value of $h_s^-$ and that catalytic efficiency saturates at large values of $h_s^-$. {\bf C.} Catalytic efficiency as a function of inverse temperature $\b$, showing exponential scaling. The exponent is $1-a$ where $a$ is given in Eq.~\eqref{eq:a}. See Fig.~\ref{fig:second} for a representation of the mean first-passage times from which $\eta_{C+S\to C+2P}$ is computed, along with standard deviations based on the second moment of these first-passage times.
 \label{fig:para}}
\end{center} 
\end{figure}

Using our approach to exactly solve the dynamics for our lattice model, we obtain $\eta_{C+S\to C+2P}=T_{S\to 2P}/T_{C+S\to C+2P}$ for a putative catalyst consisting of two binding sites, $\S=\{1,3\}$ with a common interaction energy $\e_{cs}$. The results lead to several observations~\cite{Rivoire.2023} (Fig.~\ref{fig:para}). First, catalysis ($\eta_{C+S\to C+2P}>1$) occurs only for sufficiently small values of the interaction strength $\e_{cs}$ and sufficiently large values of the reverse activation barrier $h_s^-$. Second, catalysis is optimal when $\e_{cs}$ takes an intermediate value and when $h_s^-$ is maximal. Finally, $\eta_{C+S\to C+2P}$ scales exponentially with $\b$ for large $\b$, with an exponent that can be determined analytically (see Appendix A). As we will see, this exponent quantifies the catalytic efficiency of different mechanisms independently of their geometric peculiarities. It has a simple physical interpretation in terms of activation energy. For example, if we consider the spontaneous reaction $S\to 2P$ with activation energy $h_s^+$, such that for large $\b$, $T_{C+S\to C+2P}\sim e^{\b h_s^+}$, we generally expect $T_{C+S\to C+2P}\sim e^{\b a h_s^+}$, where $a<1$ represents the factor by which the catalyst reduces the activation energy. The smaller $a$, the more efficient the catalysis, with $a=0$ representing perfect or diffusion-limited catalysis, where energetic barriers are effectively eliminated. Given such a scaling, $\eta_{C+S\to C+2P}\sim e^{(1-a)h_s^+}$.

\section{Results}

\subsection{One-way catalysis}

Is a catalyst for one reaction necessarily a catalyst for the reverse reaction? As such, the question may be considered ill-posed, since catalysis is not only an intrinsic property of the catalyst, but depends on the extrinsic conditions under which it operates, which include, for example, the volume and the temperature~\cite{Sakref.2023}. However, we can ask whether a molecule can act as a catalyst for a reaction under certain extrinsic conditions, while never acting as a catalyst for the reverse reaction under any extrinsic conditions. 

The reverse reaction for the dissociation reaction $S\to 2P$ is the association reaction $2P\to S$, whose substrates are now the two free particles (which we still denote as $2P$) and whose product is a bound dimer (which we still denote as $S$). As with the forward reaction $S\to 2P$, we can ask whether a particular configuration of binding sites provides catalysis for the reverse reaction $2P\to S$, using $\eta_{C+2P\to C+S}>1$ as a criterion. The exact same approach based on first-passage times applies, except that the initial and final states are reversed. The key difference is in the mechanisms: the forward reaction $S\to 2P$ involves a purely energetic activation barrier $h_s^+$, while the reverse reaction $2P\to S$ involves, in addition to an energetic activation barrier $h_s^-$, an entropic barrier due to the need for two particles to diffuse to find each other. As we show, some configurations of the binding sites can only lower this entropic barrier and not any energetic barrier, in which case the reverse reaction $2P\to S$ is catalyzed but not the forward reaction $S\to 2P$.

\begin{figure*}[t]
\begin{center}
\includegraphics[width=.95\textwidth]{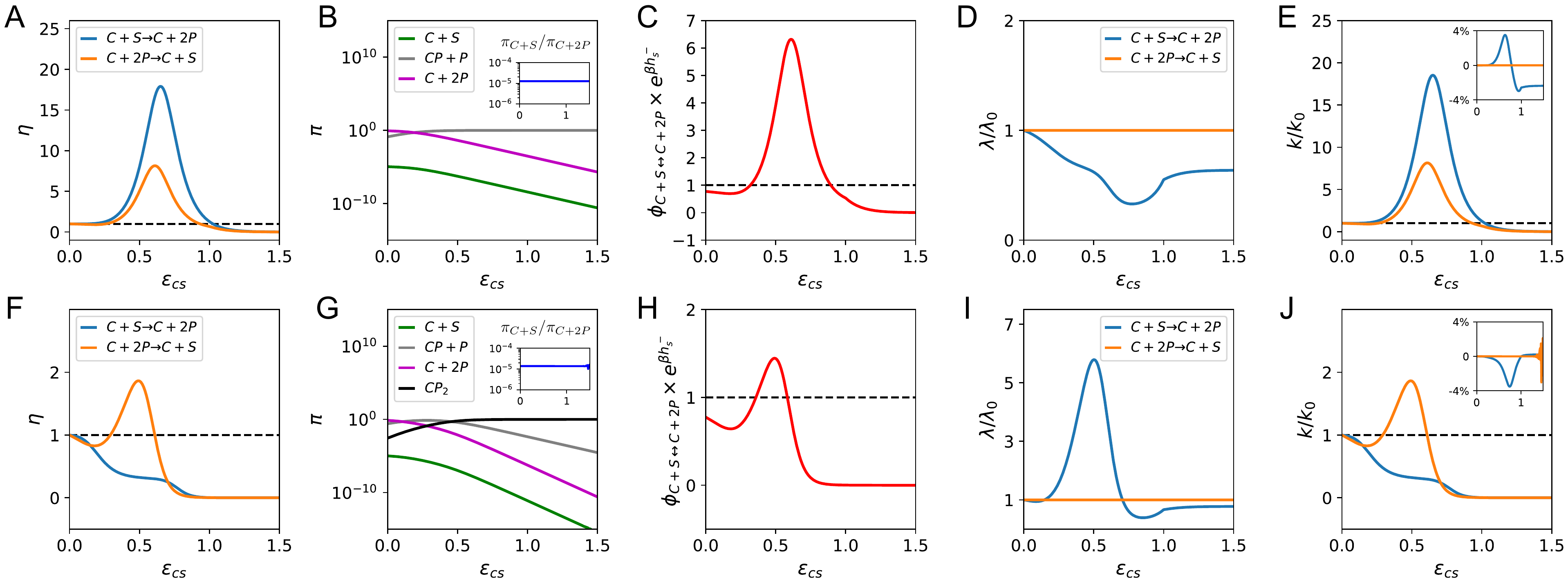}
\caption{Forward versus reverse catalysis with a two-site catalyst, with two binding sites at distance $d=2$ (top row), or four-site catalyst, with four consecutive binding sites (bottom row). {\bf A.}~$\eta_{C+S\to C+2P}$ (in blue) versus $\eta_{C+2P\to C+S}$ (in orange), showing comparable behavior as a function of interaction strength $\e_{cs}$. {\bf B.} Steady-state probabilities of the three macrostates $C+S$, $C+2P$, and $CP+P$, showing a constant ratio $\pi_{C+S}/\pi_{C+2P}$ (inset), consistent with an unchanged equilibrium constant. {\bf C.}~Steady-state flux $\phi_{C+S\leftrightarrow C+2P}$ at which transitions occur between $C+S$ and $C+2P$ (rescaled by the constant factor $e^{\b h_s^-}$ to be of order one). {\bf D.} Steady-state probabilities $\l_{C+S\to C+2P}$ (in blue) and $\l_{C+2P\to C+S}$ (in orange) for the system to be on its way from $C+S$ to $C+2P$ or from $C+2P$ to $C+S$, normalized to their values $\l_0$ when $\e_{cs}=0$. Although not visible in this representation of relative variations, both quantities vary by the same absolute amount, since $\l_{C+S\to C+2P}+\l_{C+2P\to C+S}=1$ for any value of $\e_{cs}$. {\bf E.} Catalytic rates $k_{C+S\to C+2P}=\phi_{C+S\to C+2P}/\l_{C+S\to C+2P}$ (in blue) and $k_{C+2P\to C+S}=\phi_{C+2P\to C+S}/\l_{C+2P\to C+S}$ (in orange). Normalized to their values $k_0$ when $\e_{cs}=0$, these rates coincide within a few percent with $\eta_{C+S\to C+2P}$ and $\eta_{C+2P\to C+S}$ (relative difference shown in inset). {\bf F-G} Comparable plots for a system with four consecutive binding sites. The main difference is that catalysis is only observed in the opposite direction, with $\eta_{C+S\to C+2P}<1$ for all values of $\e_{cs}$ (panel F).
 \label{fig:rev}}
\end{center} 
\end{figure*}

From solving numerically the Master equation for our lattice model, we find that the design considered so far, consisting of two binding sites at a distance of $d=2$, shows bidirectionality for most binding energies. It catalyzes dissociation $S\to 2P$ and reverse dimerization $2P\to S$ over nearly the same range of interaction strengths $\e_{cs}$ (Fig.~\ref{fig:rev}A), although it is possible to find values of $\e_{cs}$, e.g. $\e_{cs}=1$, for which (weak) forward catalysis occurs $(\eta_{C+S\to C+2P}>1)$ but not reverse catalysis ($\eta_{C+2P\to C+S}<1$). Changing the geometry of the binding sites to form four consecutive binding sites, $\S=\{1,2,3,4\}$ with the sites labeled as in Fig.~\ref{fig:scheme}A, provides an example of strictly one-way catalysis. As shown in Fig.~\ref{fig:rev}F, forward catalysis is never observed ($\eta_{C+S\to C+2P}<1$ for any $\e_{cs}$), while reverse catalysis is observed ($\eta_{C+2P\to C+S}>1$ for some $\e_{cs}$). 

As anticipated, the difference is due to the different nature of the forward and backward kinetic barriers of the spontaneous reaction. The forward reaction $S\to 2P$ is limited by a purely energetic barrier, while the backward reaction $2P\to S$ includes an entropic barrier associated with the time it takes for the two particles to diffuse toward each other. A patch of four binding sites only reduces this entropic barrier by increasing the chance of two particles interacting with each other, and therefore acts only in one direction.

More precisely, to accelerate the forward reaction $S\to 2P$, a catalyst must specifically stabilize the transition state $S^\dd$ of the reaction, when the two particles are at a distance $d=2$ (Fig.~\ref{fig:para}B). This is achieved with two binding sites separated by exactly the distance between the particles in the transition state: the total interaction energy is $2\e_{cs}$ for the transition state complex $C\:S^\dd$, while it is only $\e_{cs}$ for $CS$ and $CP$ (Fig.~\ref{fig:para}C). This mechanism of ``strain catalysis''~\cite{Haldane1930} follows Pauling principle~\cite{PAULING.1948}, with a geometry complementary to the transition state. Now, in contrast, the four-site design does not discriminate the substrate $S$, the transition state $S^\dd$, and the product $2P$: all bind with the same total interaction energy $2\e_{cs}$. On the other hand, the reverse reaction $2P\to S$ has both an energetic barrier, represented by $h_s^-$ (Fig.~\ref{fig:para}B), and an entropic barrier, corresponding to the need for the two particles to diffuse toward each other. While a patch of four binding sites cannot lower the energetic barrier, it can lower the entropic barrier by keeping the two particles in close proximity until they cross the energetic barrier. This mechanism of catalysis by proximity does not require any special geometry but applies only to multi-reactant reactions and is therefore only effective for the reverse reaction $2P\to S$.

\subsection{Catalysis at thermal equilibrium}

How is the observation of one-way catalysis consistent with thermodynamic principles? Indeed, one-way catalysis is often erroneously ruled out on the basis that catalysis does not change the equilibrium constant and therefore must accelerate both sides of a reaction equally~\cite{vsima2015catalysis}. To clarify this point, here we analyze our models under steady-state conditions that reproduce equilibrium properties.

Chemical equilibrium is typically defined at the macroscopic level, involving a large number of well-mixed particles. In this context, it is characterized by a specific ratio of substrate to product concentrations, known as the equilibrium constant. For the reaction $S\harp{}2P$ studied so far, chemical equilibrium is thus characterized at the macroscopic level by $[S]/[P]^2=e^{h_s^+-h_s^-}$ where $[S]$ and $[P]$ denote the concentrations of substrates and products. In the single-molecule context in which we analyze our model, we can instead invoke the ergodic principle to characterize chemical equilibrium by the ratio $\pi_S/\pi_{2P}$ of the steady-state probabilities of the macrostates $S$ and $2P$, $\pi_S$ and $\pi_{2P}$, which are are straightforward to calculate for our small system (Appendix C). We verify that the presence of binding sites preserves the chemical equilibrium, $\pi_{C+S}/\pi_{C+2P}=\pi_S/\pi_{P}$, independent of the geometry of the binding sites or the interaction strength $\e_{cs}$ (Fig.~\ref{fig:rev}B,G, insets). However, the values of $\pi_{C+S}$ and $\pi_{C+2P}$ change with $\e_{cs}$ as other alternative macrostates involving interactions with the binding sites are populated (Fig.~\ref{fig:rev}B,G).

While catalysis does not affect the relative occurrence of free substrates and free products at steady state, it does affect the flux between these states, i.e.~the frequency with which two particles go from a free dimer (macrostate $C+S$) to two free particles (macrostate $C+2P$). To make this point, we draw on concepts from transition path theory~\cite{Metzner.2009}. The first concept is that of flux: in any dynamics defined by a Markov process, as is the case in our microscopic model, the steady-state flux $\phi_{A\to B}$ from one macrostate $A$ to another macrostate $B$ can be defined as the frequency with which the system first reaches a microscopic configurations in $B$ after visiting the set $A$ of microscopic configurations. Formally, for a trajectory $x(t)$, we can define the time of the last visit to a configuration in $A$ as $\ell_A(t)=\sup\{t'\leq t:x(t')\in A\}$ and similarly for $\ell_B(t)$. We say that $x(t)$ last visited $A$ if $\ell_A(t)>\ell_B(t)$; otherwise it last visited $B$. Along a trajectory, we have times where $\ell_A(t)>\ell_B(t)$ and others where $\ell_A(t)>\ell_B(t)$. If the number of times we go from from $\ell_A(t)>\ell_B(t)$ to $\ell_A(t)<\ell_B(t)$ over an interval of time $T$ is $N_{A\to B}(T)$, then $\phi_{A\to B}=\lim_{T\to\infty}N_{A\to B}(T)/T$ (see Fig.~\ref{fig:illus} for a graphical illustration). Since there must be as many transitions from $\ell_A(t)>\ell_B(t)$ to $\ell_A(t)<\ell_B(t)$ than from $\ell_B(t)>\ell_A(t)$ to $\ell_B(t)<\ell_A(t)$, the flux is symmetric with $\phi_{A\to B}=\phi_{B\to A}$, which justifies denoting it by $\phi_{A\leftrightarrow B}$. Note that this symmetry holds even if the dynamics involve microscopic configurations that are neither in $A$ nor in $B$. Applied to $A=C+S$ and $B=C+2P$, this symmetry captures the intuition that a catalyst has a symmetrical incidence on the two directions of a reaction. Fluxes can be computed exactly for our small system (Appendix C). We find that the flux $\phi_{C+S\leftrightarrow C+2P}$ depends non-trivially on the interaction strength $\e_{cs}$ with a maximum at an intermediate value of $\e_{cs}$, for both the two-site and the four-site designs (Fig.~\ref{fig:rev}C-H).

To clarify how a symmetric flux $\phi_{A\leftrightarrow B}$ between two macrostates $A$ and $B$ relates to asymmetric mean first-passage times $T_{A\to B}$ and $T_{B\to A}$, one must consider another concept from transition path theory, the steady state probabilities $\l_{A\to B}$ and $\l_{B\to A}$ that the system is in the process of transitioning from $A$ to $B$ or from $B$ to $A$~\cite{Noe.2009}. Formally, $\l_{A\to B}$ is defined as the steady-state probability that the system last visited $A$ and not $B$, and similarly for $\l_{B\to A}$. In terms of the last visiting times $\ell_A(t)$ and $\ell_B(t)$, $\l_{A\to B}$ is the probability that $\ell_A(t)>\ell_B(t)$ along a long trajectory: $\l_{A\to B}=P[\ell_A(t)>\ell_B(t)]$ (see Fig.~\ref{fig:illus} for a graphical illustration). These transition probabilities can again be computed exactly (Appendix C). $\l_{A\to B}$ and $\l_{B\to A}$ are generally different, but directly related: at any given time, the system is either transitioning from $A$ to $B$ or from $B$ to $A$, so that $\l_{A\to B}+\l_{B\to A}=1$. 
This follows from the observation that either $\ell_A(t)>\ell_B(t)$ or  $\ell_A(t)<\ell_B(t)$, so that $P[\ell_A(t)>\ell_B(t)]+P[\ell_A(t)<\ell_B(t)]=1$ and therefore $\l_{A\to B}+\l_{B\to A}=1$.
Combining the flux $\phi_{A\leftrightarrow B}$ and the transition probability $\l_{A\to B}$ leads to the rate $k_{A\to B}$ at which a system in $A$ transitions to $B$, defined by $k_{A\to B}=\phi_{A\to B}/\l_{A\to B}$~\cite{Noe.2009}. 

Since $A$ and $B$ represent two different sets of configurations, the rate $k_{A\to B}$ provides a coarse-grained description of the microscopic dynamics as a Markov process with states $A$ and $B$. In general, this description is not exact. In particular, the times spent in $A$ and $B$ are generally not exponentially distributed as expected from a Markov process. However, this approximation is typically made in the context of chemical reactions, where it is often accurate. To the extent that this approximation is valid, $k_{A\to B}$ can be interpreted as a transition rate from $A$ to $B$ and be related to the mean first-passage time $T_{A\to B}$ from $A$ to $B$ by $k_{A\to B}\simeq 1/T_{A\to B}$~\cite{Berezhkovskii.2019} (note that $k$ is then a transition rate between macrostates distinct from the microscopic transition rates between microstates denoted by $\rho$). More precisely, the mean first-passage times $T_{A\to B}$ and $T_{B\to A}$ are related to the flux $\phi_{A\leftrightarrow B}$ by $\phi_{A\to B}\simeq 1/(T_{A\to B}+T_{B\to A})$ and $\l_{A\to B}\simeq T_{A\to B}/(T_{A\to B}+T_{B\to A})$~\cite{Berezhkovskii.2019}. These general mathematical relationships show how the maintenance of equilibrium properties, which is a matter of  {\it flux} between macrostates, relates with asymmetric catalysis, which is a matter of {\it rates} between macrostates:
\begin{align}
 \phi_{C+S \leftrightarrow C+2P}& \simeq k_{C+S \to C+2P} \lambda_{C+S \to C+2P} \\
 &\simeq k_{C+2P \to C+S} \lambda_{C+2P \to C+S}
\end{align}
As above, $\simeq$ indicates an approximation that becomes a strict identity when the coarse-grained dynamics between macrostates is strictly a Markov process.

Calculating $\l_{C+S\to C+2P}$ and $\l_{C+2P\to C+S}$ for the two binding site geometries considered so far, we find that $\l_{C+S\to C+2P}$ undergoes large relative changes when $\e_{cs}$ is varied, but not $\l_{C+2P\to C+S}$ (Fig.~\ref{fig:rev}D,I). Both, however, undergo the same absolute changes since $\l_{C+S\to C+2P}+\l_{C+2P\to C+S}=1$. The difference is that $\l_{C+S\to C+2P}\gg\l_{C+2P\to C+S}$ as a consequence of the parameters chosen in Fig.~\ref{fig:rev}, which favor $C+2P$ over $C+S$ since $e^{-\b h_s^-}\ll e^{-\b h_s^+}$. Using the ratio of $\phi_{C+S\leftrightarrow C+2P}$ over $\l_{C+S\to C+2P}$ or $\l_{C+2P\to C+S}$ to estimate the rates $k_{C+S\to C+2P}$ and $k_{C+2P\to C+S}$, we verify $k_{C+S\to C+2P}/k_{S\to 2P}\simeq\eta_{C+S\to C+2P}$ and $k_{C+2P\to C+S}/k_{2P\to S}\simeq\eta_{C+2P\to C+S}$ (Fig.~\ref{fig:rev}A,E,F,J). This justifies to treat the coarse-grained dynamics between macrostates as approximately Markovian. For the models and parameters of Fig.~\ref{fig:rev}, this approximation holds within 4\% (insets of Fig.~\ref{fig:rev}E,J). Below, we generalize such coarse-grained descriptions to account for the other macrostates where the binding sites are occupied by one or two particles.

\subsection{Entropic versus strain catalysis}

The two-site and four-site geometries are two alternative designs for catalyzing the dimerization reaction $2P\to S$, using different mechanisms: the four-site system, with four consecutive binding sites, relies only on catalysis by proximity, while the two-site system, with two binding sites at distance $d=2$, involves a specific stabilization of the transition state. Here we compare the efficiency of these two mechanisms by extending the scaling analysis previously performed for the two-site catalysis of the cleavage reaction $S\to 2P$~\cite{Rivoire.2023} (see Appendix A).

Fig.~\ref{fig:scaling} shows that the two mechanisms have different scaling laws in the low temperature limit. With the two-site system we recover the conclusions obtained with the forward reaction $S\to 2P$~\cite{Rivoire.2023}: the optimal interaction strength scales with the activation energy, here $h_s^-$, as $\e_{cs}=h_s^-/2$, leading to an optimal catalytic efficiency $\eta_{C+2P\to C+S}$ that increases with $h_s^-$ as $\eta_{C+2P\to C+S}\sim e^{\b h_s^-/2}$ (see Appendix A). As for the forward reaction, the activation energy is reduced by a factor of 2 at most. For the four-site system, however, we observe that the optimal interaction strength scales with the activation energy $h_s^-$ as $\e_{cs}\sim h_s^-/3$ and that the optimal catalytic efficiency $\eta_{C+2P\to C+S}$ reaches a plateau as $h_s^-$ increases (Fig.~\ref{fig:rev}D-F). 

\begin{figure}[t]
\begin{center}
\includegraphics[width=\linewidth]{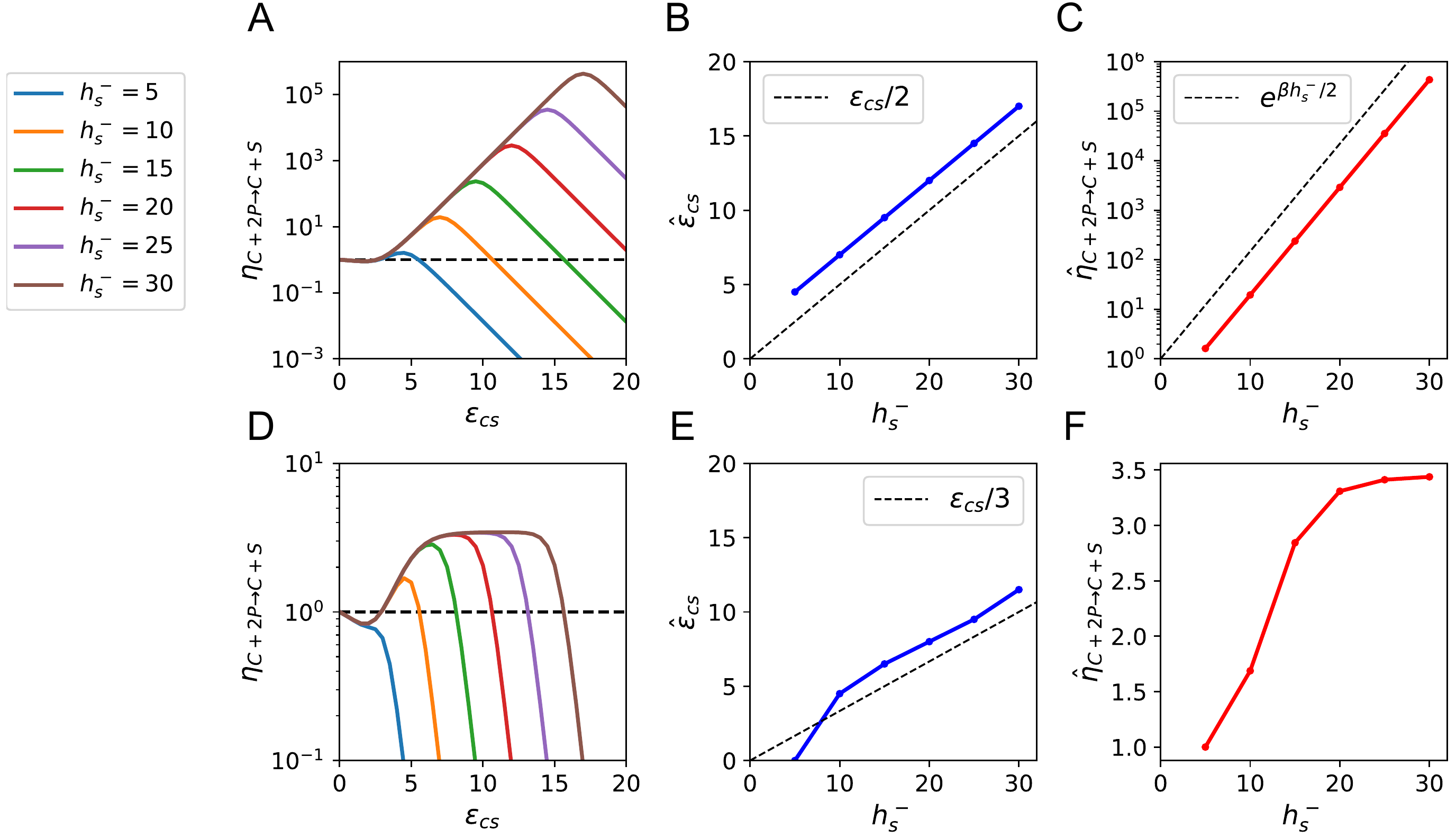}
\caption{Low temperature scaling laws for the catalysis of the dimerization reaction $2P\to S$ for a two-site (top row) or a four-site (bottom row) catalyst. {\bf A.} Catalytic efficiency $\eta_{C+2P\to C+S}$ as a function of interaction strength $\e_{cs}$ for different activation barriers $h_s^-$, showing the shape of a Volcano plot as for the opposite reaction $S\to 2P$ (Fig.~\ref{fig:para}A). {\bf B.} Scaling of the optimal interaction strength $\e_{cs}$ as a function of $h_s^-$. The dashed line has a slope of $1/2$, indicating a scaling of $\e_{cs}\sim h_s^-/2$. {\bf C.}~Scaling of the optimum of $\eta_{C+2P\to C+S}$ when varying $\e_{cs}$ as a function of $h_s^-$, showing a scaling $\eta_{C+2P\to C+S}\sim e^{\b h_s^-}$. {\bf D.} As in A, but for four consecutive binding sites (note the different scale on the y-axis). With four consecutive binding sites, {\bf E.} the optimal interaction strength $\e_{cs}$ scales as $h_s^-/3$ and {\bf F.} the optimal catalytic efficiency saturates at large values of $h_s^-$. 
 \label{fig:scaling}}
\end{center} 
\end{figure}

The scaling laws for the catalysis of the dimerization reaction $2P\to S$ by the two-site system are explained by the same arguments that we used for the catalysis of the dissociation reaction $S\to 2P$ (see Appendix A). These arguments are based on the assumption that the diffusion is effectively instantaneous at low temperature, where crossing the highest energy barrier is the limiting process. However, this assumption cannot be made when considering the catalysis of $2P\to S$ by four consecutive binding sites, since the time it takes for two particles to diffuse toward each other is precisely what the four-site catalyst reduces. 

To understand what limits catalysis in this case, we must consider the total area $V=L^2$ of the lattice as a parameter, since this is the parameter that controls the time spent diffusing. The coarse-grained model  (see Appendix B) that captures the phenomenology in the $h_s^+\gg h_s^-$ limit is of the form
\beq
C+2P\harp[k_{cs}]{k_d}CP+P\harp[k_{cs}]{k_d}CP_2\xa{k_s}CS\xa{k_{cs}^2} C+S
\eeq
where $k_d$ is a diffusion constant that scales with volume as $k_d\sim 1/V$, $k_{cs}$ is a dissociation rate that scales as $k_{cs}\sim e^{-\b \e_{cs}}$, and $k_s$ is the rate at which the activation barrier $h_s^-$ is crossed when the two reactants are close together that scales as $k_s\sim e^{-\b h_s^-}$. Neglecting the spontaneous reaction, we can apply the general formula for mean first-passage times of one-dimensional Markov chains (Appendix C) to obtain
\beq\label{eq:f}
T_{C+2P\to C+S}=\frac{2}{k_d}+\frac{1}{k_s}+\frac{1}{k_{cs}^2}+\frac{k_{cs}}{k_d^2}+\frac{k_{cs}}{k_dk_s}+\frac{k^2_{cs}}{k_d^2k_s}.
\eeq
In the low temperature limit, where $k_s$ is small, the optimal value of $k_{cs}$ is determined by $1/k_{cs}^2+k_{cs}/k_dk_s$, whose maximum is obtained when $k_{cs}\sim k_s^{1/3}$; this explains the scaling of the optimal $\e_{cs}$ as $h_s^-/3$ in Fig.~\ref{fig:scaling}E. In the same limit, however, the value of $T_{C+2P\to C+S}$ is dominated by $2/k_d+1/k_s$, which is independent of $\e_{cs}$; this explains the plateau in Fig.~\ref{fig:scaling}F and has a simple interpretation: catalysis by promixity is limited by the time for the reactants to find the catalyst ($2/k_d$) and the time for them to cross the activation energy when in proximity ($1/k_{s}$), which are independent of $\e_{cs}$.

Despite operating on different principles and exhibiting different scaling behaviors, the two-site and four-site systems both give rise to Volcano plots with an optimal interaction energy at a finite value (Fig.~\ref{fig:scaling}A,D), in accordance with the Sabatier principle~\cite{Medford.2015}. However, the shape of these Volcano plots is different. In the case of the four-site system operating by catalysis by proximity, a plateau is observed that can be understood from the coarse-grained model: this plateau extends to $\sim h_s^-/2$, which marks a crossover between the dominance of the term $1/k_s$ and the term $1/k_{cs}^2$ in Eq.~\eqref{eq:f}, occurring when $k_{cs}\sim k_s^{1/2}$. In summary, optimal catalysis by proximity require neither a fine-tuned geometry nor a fine-tuned interaction strength, unlike optimal strain catalysis. However, catalysis by proximity is generally much less efficient due to the fact that the entropic barrier that it reduces is typically lower than the energetic barrier that strain catalysis reduces.

\subsection{Optimal rigid catalysis} 

We have so far considered two geometries of the binding sites and found that for the forward reaction $S\to 2P$ only the two-site geometry, where the binding sites are separated by $d=2$, can confer catalysis. In previous work~\cite{Rivoire.2023}, we showed that a single binding site cannot confer catalysis, nor can two sites at a distance different from the transition state geometry $d=2$. We also showed that introducing asymmetry into the two-site system by allowing each binding site to have a different interaction energy does not provide any improvement. However, we can still wonder if a system with more than two sites, possibly with different interaction energies, can outperform a symmetric two-site system, i.e., lead to higher values of $\eta_{C+S\to C+2P}$. 

\begin{figure}[t]
\begin{center}
\includegraphics[width=.9\linewidth]{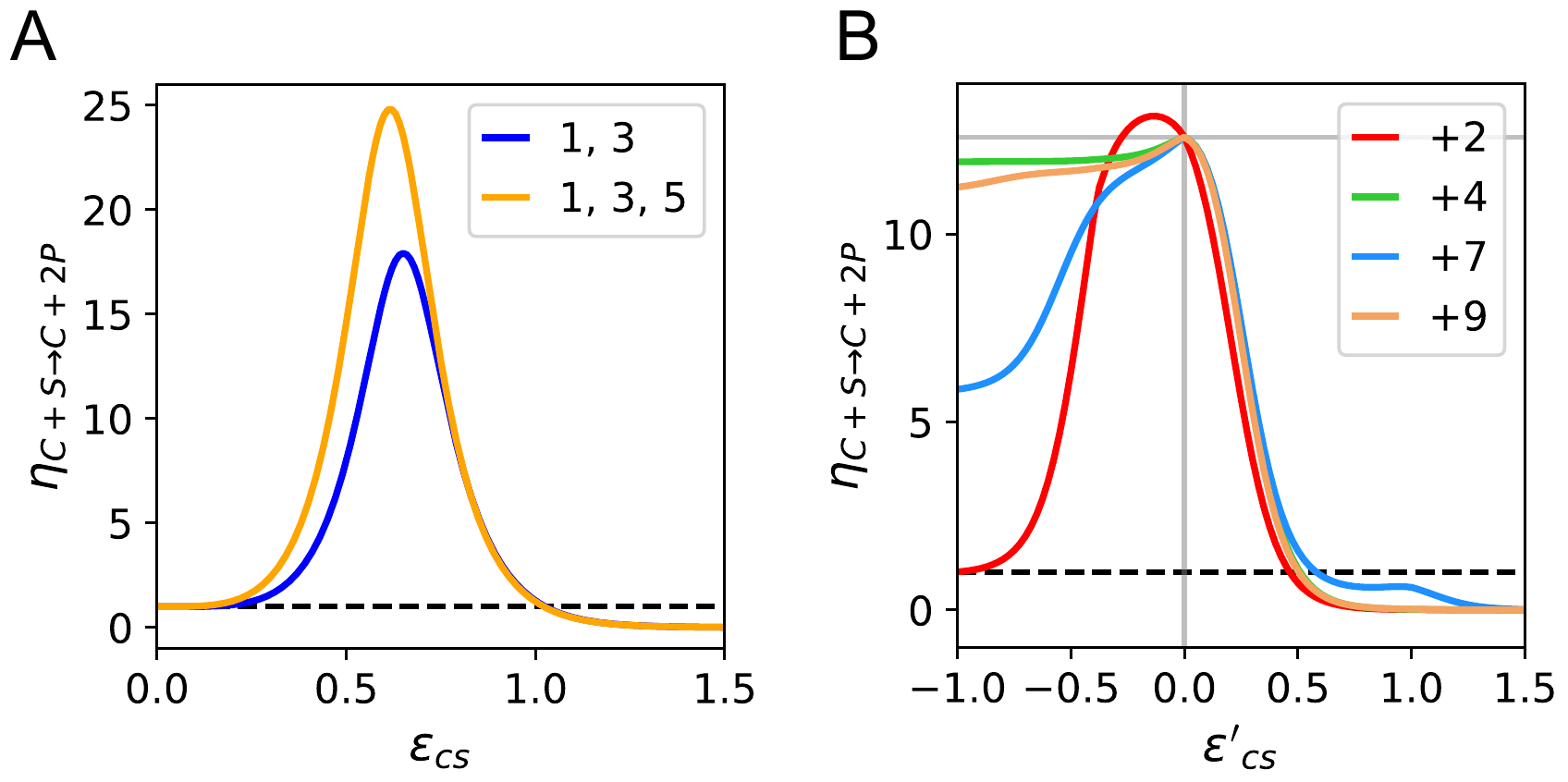}
\caption{Catalysis of the cleavage reaction $S\to 2P$ when a third binding site is introduced in addition to the two red binding sites shown in Fig.~\ref{fig:scheme}. {\bf A.} Adding another site at $j=5$ effectively increases the number of site pairs at distance $d=2$ and thus provides an entropic improvement. For a fair comparison, $\eta_{C+S\to C+EP}$ is estimated by excluding from $C+S$ and $C+2P$ all sites at the bottom of the lattice, which explains the slight difference of the blue curve from that of Fig.~\ref{fig:para}A. {\bf B.} Addition of a third site at different positions next to one of the two binding sites at $j=1$ and $j=3$ (Fig.~\ref{fig:scheme}A). The labels indicate the third location where the binding site is added. Only when a third site is added between the other two (at $j=2$) is a small improvement observed, for a destabilizing energy $\e_{cs}'<0$.
 \label{fig:more}}
\end{center} 
\end{figure}

One possibility is to have more pairs of sites that implement the same mechanism of strain catalysis. For example, in the context of our $6\times 6$ lattice, we can add a third binding site at $j=5$ (Fig.~\ref{fig:scheme}A), which, given the periodic boundary conditions, provides 3 pairs of binding sites at distance $d=2$. This model can be considered a model of heterogeneous catalysis, which typically involves a surface of periodically repeated binding sites. We verify that the additional binding site does provide an improvement (Fig.~\ref{fig:more}A), but this corresponds to an entropic factor that leaves the low-temperature scaling properties unchanged, in particular the factor $a$ that controls the scaling of the catalytic efficiency $\eta_{C+S\to C+2P}$ with the amplitude of the activation energy (see Eq.~\eqref{eq:a}). 

Another question is whether it is possible to improve the two-site catalyst by adding a third binding site through a new or additional mechanism. As a systematic approach, we consider the addition of a third binding site at any site adjacent to one of the two binding sites at distance $d=2$. The binding energy at this new site is taken to be $\e_{cs}'$, possibly different from the binding energy $\e_{cs}$ at the two sites $j=1$ and $j=3$ (Fig.~\ref{fig:scheme}A). Considering the symmetries, four geometries are possible where the third binding site is located at sites $j=2$, 4, 7 or 9 (Fig.~\ref{fig:scheme}A). We find that only in one case, when the third site is between the other two ($j=2$), a small improvement is observed, provided that the interaction is repulsive, i.e. $\e_{cs}'<0$ (Fig.~\ref{fig:more}B). 

First, we can understand why adding a third site at any other position is detrimental. Naively, a third site at $j=4$, for example, can be thought of as favoring the release step $C\.P+P\to C+2P$ by providing a ``stepping stone'' that replaces the crossing of a barrier $\e_{cs}$ with the crossing of two smaller barriers whose sum is $\e_{cs}$. Strain catalysis with two binding sites at distance $d=2$ essentially works like this: the barrier $h_s^+$ is replaced by two smaller barriers $h_s^+-\e_{cs}$ and $\e_{cs}$ (see Eq.~\eqref{eq:3terms} with $h_s^+\ll h_s^-$). But, as we noted, this mechanism only works in the presence of a reverse barrier $h_s^->0$. Simply replacing one barrier by two smaller ones in the absence of a reverse barrier never results in acceleration. Mathematically, using the formula for mean first-passage times of one-dimensional Markov processes (Appendix C) this amounts to replacing $T_{1\to 2}=e^{\b \e_{12}}$ by $T_{1\to 3\to 2}=e^{\b \e_{13}}+e^{\b \e_{32}}+e^{\b (\e_{13}+\e_{32}-e_{32})}$ with the constraint $\e_{13}+\e_{32}=\e_{12}$, which implies $T_{1\to 3\to 2}\geq T_{1\to 2}$.

The same reasoning suggests that no benefit should be expected from a third binding site between the other two: introducing a ``stepping stone'' of intermediate energy cannot accelerate the transition from $C\.S\to C\:S^\dd$. But if such a naive mechanism of ``substrate destabilization''~\cite{Jencks.1993b18} cannot be effective, how to explain the (small) improvement observed in Fig.~\ref{fig:scheme}B? Denoting by $C\:S$ the new state where $S$ is bound to two consecutive sites ($(i_1,i_2)=(1,2)$ or $( 2,3)$), the catalytic cycle in the presence of a third middle binding site at $j=2$, is described at a coarse-grained level by a Markov chain of the form
\beq
C+S\harp[k_{-1}]{k_1}C\.S\harp[k_{-2}]{k_2}C\: S \harp[k_{-3}]{k_3} C\:S^\dd\harp[k_{-4}]{k_4}C\.P+P\harp[k_{-5}]{k_5}C+2P
\eeq
If we first assume that the third site has zero binding energy ($\e_{cs}'=0$), we have $k_2=k_{-2}=k_{-3}=k_4=1$ and the optimal interaction energy $\e_{cs}=h_s^+/2$ for large $\b$ implies $k_3=k_5$, leaving five equally dominant terms
\beq
T_{C+S\to C+2P}=\frac{1}{k_3}+\frac{1}{k_5}+\frac{k_{-2}}{k_2k_3}+\frac{k_{-3}}{k_3k_4}+\frac{k_{-2}k_{-3}}{k_2k_3k_4}\simeq 5e^{h^+/2}
\eeq
By introducing a middle site with non-zero binding energy, the rate $k_{-3}$ is changed to $k_{-3}e^{-\e_{cs}'}$ and with $\e_{cs}'<0$ the two terms involving $k_{-3}$ are no longer dominant, leading to $T_{C+S\to C+2P}\simeq 3e^{h_s^+/2}$. The factor $3/5$ is the gain observed in Fig.~\ref{fig:scheme}B. Physically, it is an entropic factor corresponding to a favored transition $C\.S\to C\:S^\dd$, achieved by promoting a proper alignment of the substrate relative to the binding sites. As an entropic factor, it provides only a small improvement and has no effect on the low-temperature scaling behavior. In particular, $a$ is still given by Eq.~\eqref{eq:astar}, and the three-site catalyst is therefore subject to the same $a\geq 1/2$ limitation as the two-site catalyst~\cite{Rivoire.2023}. 

\section{Conclusion}

Fundamental thermodynamic principles dictate that catalysis, the acceleration of a reaction without any energy input or alteration of the catalyst, must preserve equilibrium properties. At first glance, this seems to exclude one-way catalysis where only one direction of a reaction is accelerated. However, catalysis is usually studied under non-equilibrium conditions, starting from a high concentration of substrate to study the catalysis of the forward reaction, or from a high concentration of product to study the catalysis of the backward reaction. Here we revisited the relationship between directional catalysis and chemical equilibrium by analyzing single microscopic trajectories, where both concepts are relevant. Using notions from transition path theory~\cite{Metzner.2009}, we showed in the context of a simple solvable model how the two concepts are encoded in two distinct but related mathematical quantities: transformation {\it rates}, which are asymmetric with respect to the two endpoints, and steady-state {\it fluxes}, which are symmetric at chemical equilibrium. In essence, catalysis pertains to rates, while equilibrium pertains to fluxes.  

We also provided a simple example of a strictly one-way catalyst, which can possibly accelerate only one direction of a reaction. This example relies on two features of the reaction and the catalyst. First, we considered a dimerization reaction $S\harp{} 2P$ in which the kinetic barriers limiting the forward and reverse directions are of different nature, namely purely energetic for dissociation and a mixture of energetic and entropic for association. Second, we studied a catalytic mechanism by proximity involving a patch of successive binding sites, which can possibly reduce only the entropic component of a reaction barrier. Since only the reverse direction of the reaction $S\harp{} 2P$ contains an entropic component, this mechanism can only accelerate this reverse direction. However, the fact that a catalyst is more efficient at catalyzing one direction of a reaction is very general and indeed generic, as illustrated by the other catalytic mechanism we studied, based on strain. 

The distinction between two catalytic mechanisms, by proximity and by strain, led us to identify different scaling laws in the limit of large activation energies or, equivalently, small temperatures. However, we showed that for a given mechanism, the scaling of a catalyst with optimized interaction energies is robust to variations in its geometry. In particular, strain catalysis in our model can lower the activation barrier by a factor of at most two ($a\geq 1/2$), even when introducing more binding sites than the two required for minimal implementation of this mechanism. The additional sites can provide a small entropic improvement but do not change the optimal scaling. This limitation of strain catalysis is general and holds whenever the binding energies of the three reaction states -- substrate $S$, transition state $S^\dd$, and product(s) $2P$ -- are constrained to be correlated, which is generally caused by their chemical similarity~\cite{Rivoire.2024}. In our specific model, the binding energy to the transition state is $2\e_{cs}$, which cannot be increased without also increasing the binding energies to the substrate and product, both of which are $\e_{cs}$. If there is a different relationship between these three binding energies, the optimal scaling factor $a$ can differ from $a=1/2$, but a finite value $a>0$ is expected as long as no internal degrees of freedom are introduced. As previously shown within the same modeling framework, this scaling can indeed be broken by replacing fixed binding sites with flexible ones, thus implementing a form of allostery commonly observed in enzymes~\cite{Rivoire.2023}.

In summary, we have elaborated the simple framework of an exactly solvable lattice model of catalysis~\cite{Rivoire.2023} to show how it can illuminate general questions related to catalysis: (1) We have extended the analysis to cover both a forward reaction and its reverse, and to consider different geometries of binding sites constituting a rigid catalyst. In all cases, we recover Sabatier principle that an intermediate binding energy is optimal~\cite{Medford.2015}, as well as the observation that rigidity imposes a limit on the efficiency of catalysis. However, we find that this limitation can take different forms, as quantified by the factor $a$ by which catalysis lowers the activation barrier in the low-temperature limit. (2) We showed how this factor $a$ is robust to details of the geometry of catalysis, thus justifying its definition as a measure of catalytic efficiency for a class of catalytic mechanisms. (3)~We provided an example of a strictly one-way catalyst along with an explanation for its adirectionality by distinguishing two types of catalytic mechanisms, by strain or by proximity. (4) We illustrated and explained how another putative mechanism of substrate destabilization could not be effective with fixed binding sites (as opposed to an allosteric mechanism involving flexibility~\cite{Rivoire.2023}). (5) We introduced concepts from transition path theory~\cite{Metzner.2009} to coarse-grain microscopic trajectories as transitions between macrostates, and used this approach to explain how catalysis can be directional and detectable by inspecting the trajectory of single molecules, and yet maintain the equilibrium concentration at a coarse-grained, macroscopic level.

Our modeling framework could be further generalized to address other questions related to catalysis. In theoretical treatments involving catalysis in the statistical physics literature, catalytic mechanisms are almost always abstracted and treated only at the kinetic level. This is the case even when the focus is on studying specific enzymes, such as molecular motors, and when the models are intended to be mechanistic~\cite{omabegho2021allosteric}. As shown here and in previous work~\cite{Rivoire.2020,Munoz-Basagoiti.2023,Rivoire.2023,sakref2024design}, catalytic mechanisms can be easily incorporated into simple and solvable physical models, allowing one to derive scaling laws and conceptual insights in a transparent manner that does not require approximations.\\ \ \\

{\bf Acknowledgments -- } We acknowledge funding from ANR-22-CE06-0037.\\

\appendix

\section{Low-temperature limit}\label{sec:scaling}

The exponential scaling with $\b$ of the catalytic efficiency $\eta_{C+S\to C+2P}$ illustrated in  Fig.~\ref{fig:para}C is consistent with Arrhenius law. For the spontaneous reaction we indeed have $T_{S\to 2P}\sim e^{\b h_s^+}$, where the symbol $\sim$ stands for $\lim_{\b\to\infty}(\ln T_{S\to 2P})/\b=h_s^+$. In the presence of catalysis, we can expect a similar scaling $T_{C+S\to C+2P}\sim e^{\b ah_s^+}$ but with a factor $a<1$ representing the reduction of the activation energy by the catalyst. More formally, we define this activation energy reduction by
\beq\label{eq:a}
a=\lim_{\b\to\infty}\frac{\ln T_{C+S\to C+2P}}{\ln T_{S\to 2P}}.
\eeq
With respect to this factor, $\eta_{C+S\to C+2P}\sim e^{\b(1-a)h_s^+}$. 

$a$ can be calculated analytically by noting that, at low temperatures (large $\b$), diffusion is much faster than barrier crossing. If we treat diffusion as instantaneous, or more precisely as occurring on the fastest time scale of the dynamics, which according to Eq.~\eqref{eq:rho} is $1/\rho_0=1$, all configurations of the same energy connected by diffusion are effectively equivalent. For the model presented in Fig.~\ref{fig:scheme}, we have only five different macrostates when ignoring the spontaneous reaction, $C+S$, $C\.S$, $C\:S^\dd$, $C\.P+P$ and $C+2P$ in this approximation (Fig.~\ref{fig:scheme}C), with respective energies $h_s^--h_s^+$, $h_s^--h_s^+-\e_{cs}$, $h_s^--2\e_{cs}$, $-\e_{cs}$ and $0$, and the dynamics is reduced to a one-dimensional Markov chain between these macrostates, of the form 
\beq\label{eq:1dMM}
C+S\harp[k_{-1}]{k_1}C\.S\harp[k_{-2}]{k_2}C\:S^\dd\harp[k_{-3}]{k_3}C\.P+P\harp[k_{-4}]{k_4}C+2P
\eeq
We can then apply an analytical formula for the mean first passage times of one-dimensional Markov chains (see Appendix C)
\begin{align}\label{eq:Trho}
T_{C+S\to C+2P}=&
\frac{1}{k_1}+\frac{1}{k_2}+\frac{1}{k_3}+\frac{1}{k_4}
+\frac{k_{-1}}{k_1k_2}+\frac{k_{-2}}{k_2k_3}+\frac{k_{-3}}{k_3k_4}\nonumber\\
&+\frac{k_{-1}k_{-2}}{k_1k_2k_3}+\frac{k_{-2}k_{-3}}{k_2k_3k_4}
+\frac{k_{-1}k_{-2}k_{-3}}{k_1k_2k_3k_4}
\end{align}

If we consider $\e_{cs}<h_s^+$, which is required for catalysis (otherwise release involves an activation energy $\e_{cs}$ greater than the activation energy of the spontaneous reaction $h_s^+$), and assume for simplicity $\e_{cs}<h_s^-$, an assumption which we show below to be nonrestrictive for optimal catalysis, we have
\begin{align}
&k_{-1}=e^{-\b e_{cs}},\quad 
k_2=e^{-\b (h_s^+-e_{cs})},\nonumber\\
&k_{-3}=e^{-\b (h_s^--e_{cs})},\quad 
k_4=e^{-\b e_{cs}},\\
&k_1=k_{-2}=k_3=k_{-4}=1.\nonumber
\end{align}
These expressions result in
\begin{align}
T_{C+S\to C+2P}=&
2+2e^{\b(h_s^+-\e_{cs})}+e^{\b \e_{cs}}+2e^{\b(h_s^+-2\e_{cs})} \\
&+e^{\b(2\e_{cs}-h_s^-)}+e^{\b (h_s^+-h_s^-+\e_{cs})}+e^{\b(h_s^+-h_s^-)}\nonumber
\end{align}
When $\b$ is large, this sum is dominated by the term(s) with the largest exponent. Noticing that some terms are larger than others regardless of the parameters, we get
\beq\label{eq:3terms}
T_{C+S\to C+2P}\simeq
2e^{\b(h_s^+-\e_{cs})}+e^{\b \e_{cs}}+e^{\b (h_s^+-h_s^-+\e_{cs})}
\eeq
The first term, with kinetic barrier $h_s^+-\e_{cs}$, is the time to cross the transition state and dissociate, which is reduced compared to the spontaneous reaction whose kinetic barrier is $h_s^+$. The second term, with kinetic barrier $\e_{cs}$, corresponds to the time to release a monomer from the catalyst. The last term, with kinetic barrier $h_s^+-h_s^-+\e_{cs}$, corresponds to an extra time due to recrossing the transition state while interacting with the catalyst.

This last term becomes irrelevant in the limit of an irreversible reaction where $h_s^-=\infty$. In this limit, there is a simple trade-off between the activation energy $h_s^+-\e_{cs}$ and the release energy $\e_{cs}$: one decreases with $\e_{cs}$ while the other increases. This trade-off, which is the essence of the Sabatier principle~\cite{Medford.2015}, implies an intermediate value for the optimal binding energy. In this limit, the optimum is reached when the two kinetic barriers in the trade-off are equal, leading to an optimum with $\e_{cs}=h_s^+/2$ and $T_{C+S\to C+2P}\sim 3e^{\b h_s^+/2}$. 

\begin{figure}[t]
\begin{center}
\includegraphics[width=\linewidth]{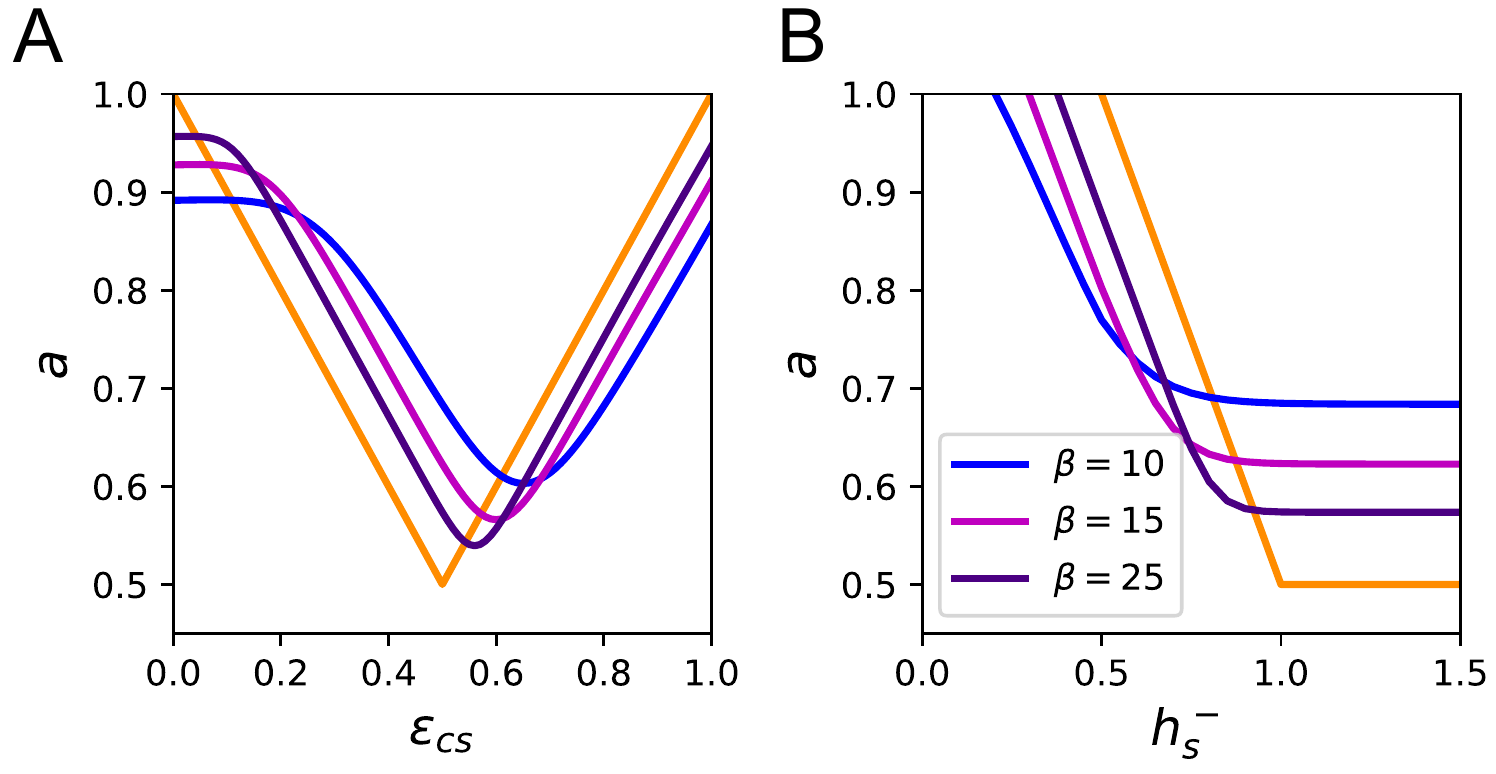}
\caption{\small Effective activation energy. The definition of the factor $a$ by which the activation energy is reduced by catalysis, given in the low-temperature limit ($\b\to\infty$) by Eq.~\eqref{eq:a}, can be extended to finite temperature as $a(\b)=(\ln T_{C+S\to C+2P})/\b$. {\bf A.} Dependence of $a(\b)$ on the interaction energy $\e_{cs}$. The orange curve reports $a=a(\beta=\infty)$ given by Eq.~\eqref{eq:astar}. {\bf B.} Dependence on the reverse activation energy $h_s^-$. These results are directly comparable to those of Fig.~\ref{fig:para}, since $\eta_{C+S\to C+2P}\sim e^{\b(1-a)h_s^+}$.
 \label{fig:a}}
\end{center} 
\end{figure}

This optimum is unchanged for large but finite values of the reverse activation energy $h_s^-$ up to the point where the third term in Eq.~\eqref{eq:3terms} comes to dominate the second term, which occurs when $h_s^-<h_s^+$. The tradeoff is then between the first and third terms, with an optimal value reached for $\e_{cs}=h_s^-/2$, in which case $T_{C+S\to C+2P}\sim 3e^{\b (h_s^+-h_s^-/2)}$. All in all, and as shown previously~\cite{Rivoire.2023}, the minimum mean first-passage time is of the form $T_{C+S\to C+2P}\sim e^{\b a^* h_s^+}$.
with
\beq\label{eq:astar}
a^*= \max(1/2,1-h_s^-/h_s^+)
\eeq
Two implications can be drawn from this calculation. First, since $a^*\geq 1/2$, this type of rigid catalysis can lower the activation barrier by at most a factor of 2. Second, catalysis requires a non-zero reverse barrier $h_s^->0$, otherwise $a=1$. While these results are derived in the $\b=\infty$ limit, they capture the dependence on the parameters $\e_{cs}$ and $h_s^-$ for finite values of $\b$ (Fig.~\ref{fig:a}).

\section{Macroscopic model}\label{sec:markov}

The representation of the catalytic cycle given by Eq.~\eqref{eq:1dMM} differs in two respects from the representation of catalytic cycles by Markov chains usually presented in the literature. First, it includes unstable macrostates: for example, if we consider $\e_{cs}<h_s^\pm$, as we did to obtain Eq.~\eqref{eq:rho}, $C\:S^\dd$ is unstable because no activation barrier separates it from the preceding and succeeding states. Second, it is derived in the limit where diffusion is instantaneous and therefore does not account for diffusion processes. One way to include them at a mean-field level is to assume mass-action kinetics. Eliminating unstable states and using mass action kinetics, Eq.~\eqref{eq:1dMM} is rewritten as
\beq\label{eq:Mark'ov}
C+S\harp[k'_{-1}]{k'_1}CS\harp[k'_{-2}]{k'_2}CP+P\harp[k'_{-3}]{k'_3}C+2P
\eeq
with
\begin{align}
& k'_1=k_d[S],\quad k'_{-1}=k_{-1},\quad k'_2=k_2/2,\nonumber\\
& k'_{-2}=k_{-3}k_d[P],\quad k'_3=k_4,\quad k'_{-3}=k_d[P]
\end{align}
where $k'_d$ is a diffusion coefficient and $[S]$ and $[P]$ are the concentrations of the dimers and monomers. With only two particles, these concentrations can take a limited number of values: $[S]=0$ or $1/V$, $[P]=0$, $1/V$ or $2/V$. The factor 2 in $k'_2=k_2/2$ follows from $1/k'_2=1/k_2+1/k_3+k_{-2}/(k_2k_3)\simeq 2/k_2$, since $k_{-2}=k_3=1\ll k_2$. 

Ignoring the spontaneous reaction, the mean first-passage time $T_{C+S\to C+2P}$ is again obtained from this coarse-grained model using the general formula applicable to one-dimensional Markov chains (Appendix C),
\beq
T_{C+S\to C+2P}=.
\frac{1}{k'_1}+\frac{1}{k'_2}+\frac{1}{k'_3}
+\frac{k'_{-1}}{k'_1k'_2}+\frac{k'_{-2}}{k'_2k'_3}
+\frac{k'_{-1}k'_{-2}}{k'_1k'_2k'_3}
\eeq
In the low temperature limit ($\b\to\infty$), this formula gives results strictly equivalent to Eq.~\eqref{eq:Trho}.

\section{Formulas relative to Markov processes}

\subsection{Transition matrix}\label{app:tm}

Given the transition rates defined in Eq.~\eqref{eq:rho}, the master equation describes the evolution of the probability $\pi(x,t)$ for the system to be in configuration $x$ at time $t$, namely
\beq
\partial_t\pi(x,t)=\sum_{y\neq x}\left(\pi(y,t)k(y\to x)-\pi(x,t)k(x\to y)\right)
\eeq
It is convenient to represent this equation in matrix form. If $L^2$ is the number of lattice sites, then the number of configurations of the two particles is $N=L^2(L^2-1)$, and the time evolution can be represented by an $N$-dimensional vector $\pi(t)$ with components $\pi_x(t)=\pi(x,t)$. The master equation can then be rewritten
\beq
\partial_t\pi(t)=Q^{\top}\pi(t)
\eeq
where $Q^\top$ is the transpose of the $N\times N$ dimensional matrix $Q$ whose components are
\begin{align}\label{eq:Q}
Q_{xy}=
\left\{
    \begin{array}{ll}
        k(x\to y) & \mbox{if } x\neq y \\
       -\sum_{y\neq x}k(y\to x)& \mbox{if } x=y.\\
    \end{array}
\right.
\end{align}
This transition matrix is known as the generator of the continuous time Markov chain. 

\begin{figure*}[t]
\begin{center}
\includegraphics[width=.9\textwidth]{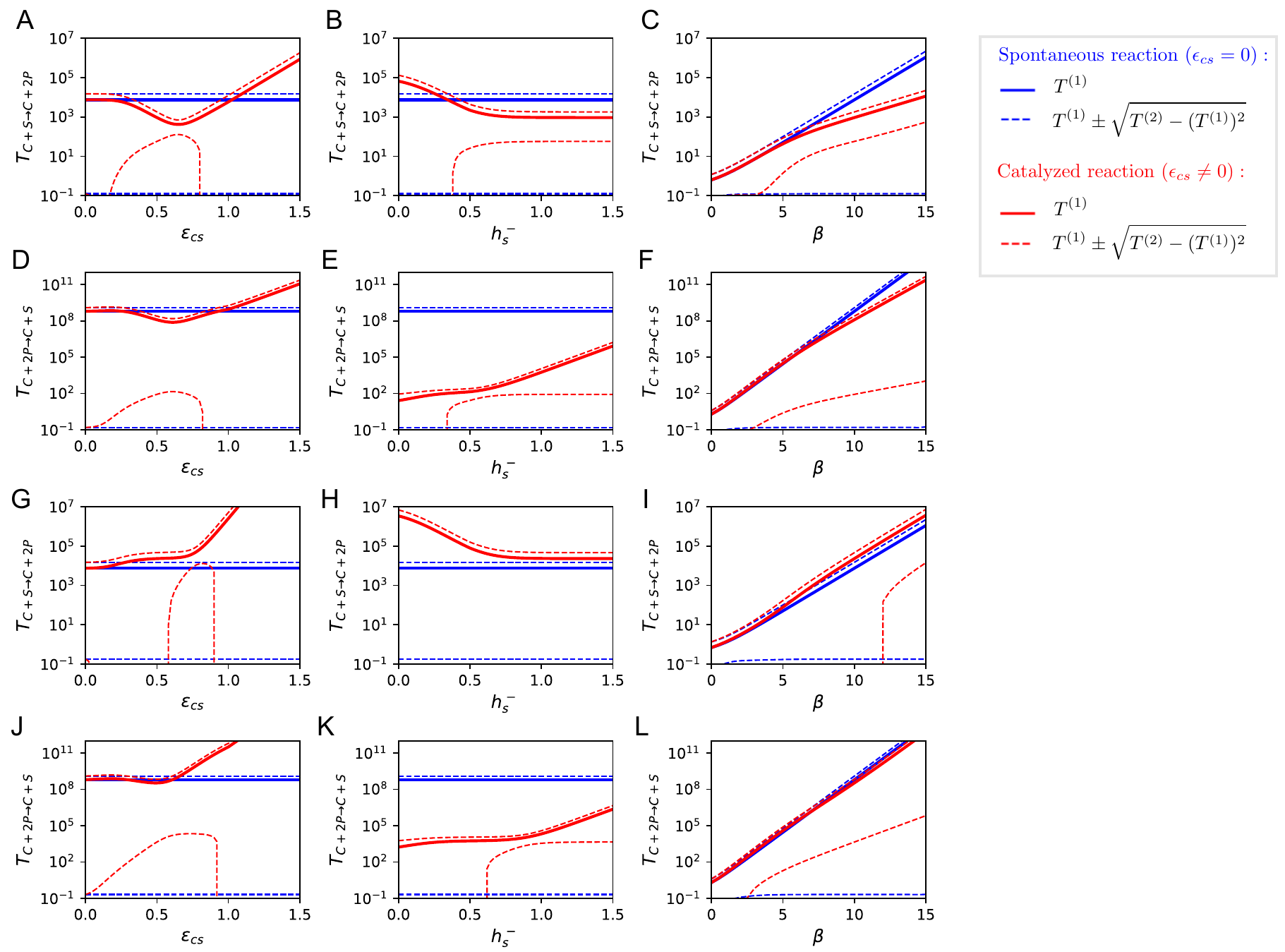}
\caption{\small Mean and standard deviation of first-passage times -- The results presented in the main text are based on the mean first-passage times $T_{C+S\to C+2P}$ or $T_{C+2P\to C+S}$. Here we show these mean times $T^{(1)}$ for the spontaneous ($\e_{cs}=0$) and catalyzed reactions ($\e_{cs}\neq 0$) as a function of different parameters and for different configurations of the binding sites defining the catalyst, along with their standard deviation, obtained from the second moment of the first-passage times $T^{(2)}$ as $\sqrt{T^{(2)}-(T^{(1)})^2}$. These standard deviations are typically of the same order as the mean first-passage times (note the log scale along the y-axis). {\bf A,B,C:} Forward reaction with two binding sites separated by a distance of 2. These plots correspond to Fig.~\ref{fig:para}, which is obtained by comparing the mean first-passage times of the spontaneous and catalyzed reactions. {\bf D,E,F:} Reverse reaction with two binding sites separated by a distance of 2. {\bf G,H,I:} Forward reaction with four consecutive sites.  {\bf J,K,L:} Reverse reaction with four consecutive sites.
 \label{fig:second}}
\end{center} 
\end{figure*}

\subsection{Steady-state probabilities}\label{app:ss}

The stationary distribution over the configurations is a $N$-dimensional vector $\pi$ that can be obtained as the left-eigenvector of $Q$ with eigenvalue $0$, satisfying $\pi^\top Q=0$, normalized to have $\sum_x\pi_x=1$. If $A$ is a set of configurations $x$, we denote by $v_A$ a $N$-dimensional vector with $(v_{A})_x=1$ if $i\in A$ and 0 otherwise. The steady-state probability to be in state $A$ is then $v_A^\top\pi$.

\subsection{Moments of first-passage times}

Let $A$, $B$ be two non-intersecting sets of configurations $x$. The mean first-passage time to reach $B$ from a configuration $x\notin B$, denoted $\tau_{x\to B}$, is solution to $\sum_{y\notin B}Q_{xy}\tau_{y\to B}=-1$.  This equation is a discrete-space analog of the backward Kolmogorov equation, also known as Dynkin equation, which involves the adjoint of the Fokker-Planck operator. The $N$-dimensional vector $\tau_{:\to B}$ is therefore given by
\begin{align}
\tau_{B\to B}&=0\nonumber\\
\tau_{\bar B\to B}&=-(Q_{\bar B,\bar B})^{-1}u_{\bar B}
\end{align}
where $u_{\bar B}$ is a vector whose dimension is the size of ${\bar B}$, with $(u_{\bar B})_x=1$ for all $x$.

The global mean first-passage time from any configuration in $A$ to $B$ is $T_{A\to B}=\sum_{x\in A}\tau_{x\to B}/|A|$, where $|A|$ is the number of elements in $A$, i.e.,
\beq
T_{A\to B}=\frac{v_A^\top \tau_{A\to B}}{v_A^\top v_A}
\eeq

The relation $\sum_{y\notin B}Q_{xy}\tau_{y\to B}=-1$, from which we obtain mean first-passage times by matrix inversion, generalizes to $\sum_{y\notin B}Q^n_{xy}\tau^{(n)}_{y\to B}=(-1)^nn$ to obtain any higher moment $n\geq 2$ of first-passage times. For each mean first-passage time $T_{A\to B}=T^{(1)}_{A\to B}$ that we compute, we can therefore also estimate its standard deviation as $\sqrt{T_{A\to B}^{(2)}-(T_{A\to B}^{(1)})^2}$, as illustrated in Fig.~\ref{fig:second}.

\subsection{Committors}

The probability $c^{B<A}_x$ for a trajectory to reach $B$ before $A$ when starting from $x$ is solution to $\sum_yQ_{xy}c^{B<A}_y=0$ if $x\in I=\overline{A\cup B}$, with boundary conditions $c^{B<A}_x=0$ if $x\in A$ and $c^{B<A}_x=1$ if $x\in B$. We therefore have $Q_{I,I}c^{B<A}_I=-\sum_{y\in B}Q_{xy}$. In short, the $N$-dimensional vector $c^{B<A}$ is given by~\cite{Noe.2009}
\begin{align}
c^{B<A}_A&=0\nonumber\\
c^{B<A}_B&=1\\
c^{B<A}_I&=-(Q_{I,I})^{-1}Q_{I,:}v_B\qquad (I=\overline{A\cup B})\nonumber
\end{align}

At steady state, $1-c^{B<A}$ also gives probability to have last been in $A$ rather than in $B$.

\begin{figure}[t]
\begin{center}
\includegraphics[width=.9\linewidth]{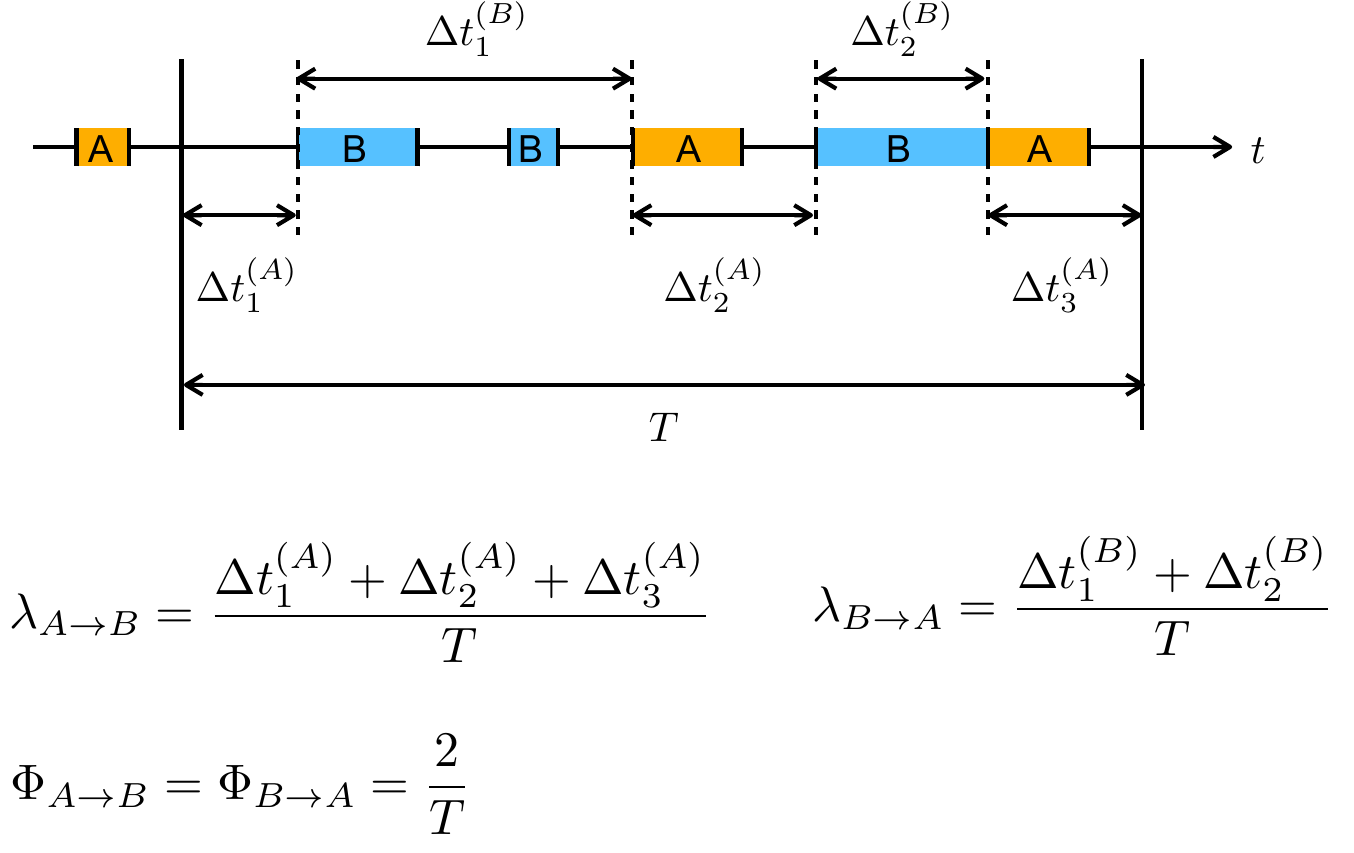}
\caption{\small Illustration of the definitions of $\l_{A\to B}$ and $\Phi_{A\to B}$ -- Over a time interval $T$, a trajectory samples different states, including states belonging to two disjoint subsets $A$ and $B$. The time spent in $A$ is shown in orange and the time spent in $B$ in blue. At any given time, the system has either last visited $A$ or last visited $B$. The durations of the time intervals when it last visited $A$ are marked as $\Delta t_i^{(A)}$ and those when it last visited $B$ are marked as $\Delta t_i^{(B)}$. $\lambda_{A\to B}$ is the fraction of time it last visited $A$ and $\lambda_{B\to A}$ the fraction of time it last visited $B$. By construction, $\lambda_{A\to B}+\lambda_{B\to A}=1$. Here the system changes twice from having last visited $A$ to having last visited $B$, so $\Phi_{A\to B}=2/T$, and also twice from having last visited $B$ to having last visited $A$, so $\Phi_{B\to A}=2/T$. The illustration is made with a finite interval $T$, but $\lambda_{A\to B}$ and $\Phi_{A\to B}$ are mathematically defined in the limit $T\to\infty$. This guarantees $\Phi_{A\to B}=\Phi_{B\to A}$, which is also noted $\Phi_{A\leftrightarrow B}$. \label{fig:illus}}
\end{center} 
\end{figure}

\subsection{Fluxes}\label{app:flux}

The flux $\phi_{A\to B}$ is the number of first times at which $B$ is reached after having visited $A$, divided by the total time. At steady state, the flux at which $A$ transitions to $B$ is the same as the flux at which $B$ transitions to $A$. It is a scalar given by~\cite{Noe.2009}
\beq
\phi_{A\leftrightarrow B}=\pi_A^\top Q_{A,\bar A}c^{B<A}_{\bar A}
\eeq\\

\subsection{Transition probabilities}\label{app:tp}

At steady-state, the probability that the system is undergoing a transition from $A$ to $B$ is the probability that is has last visited $A$ rather than $B$. It is a scalar given by~\cite{Noe.2009}
\beq
\l_{A\to B}=\pi^\top(1-c^{B<A})=1-\pi^\top c^{B<A}
\eeq

\subsection{Rates}

The rate from $A$ to $B$ is generally not the same as the rate from $B$ to $A$. It is inverse of the mean time for a trajectory to reach $B$ given that it last visited $A$. It is a scalar given by~\cite{Noe.2009}
\beq
k_{A\to B}=\frac{\phi_{A\leftrightarrow B}}{\l_{A\to B}}
\eeq

\subsection{Mean first-passage times for one-dimensional Markov chains}\label{app:1d}

The mean first passage time from $X_0$ to $X_{N+1}$ through $N$ intermediate states along a one-dimensional chain of transitions of the type
\beq
X_0\harp[k_{-1}]{k_1}X_1\harp[k_{-2}]{k_2}\dots\dots \harp[k_{-{N-1}}]{k_{N-1}}X_N\harp[k_{-N}]{k_N}X_{N+1}
\eeq
is given by
\begin{align}\label{eq:sumk}
T_c&=\sum_{j=1}^{N+1}\sum_{j=i}^{N+1}\frac{1}{k_j}\prod_{r=i}^{j-1}\frac{k_{-r}}{k_r}\\
&=\sum_{1\leq i\leq j\leq N+1}\frac{k_{-i}k_{-(i+1)}\dots k_{-(j-2)}k_{-(j-1)}}{k_{i}k_{i+1}\dots k_{j-1} k_j}.\nonumber
\end{align}
This result is for instance obtained by writing $T_c=\sum_{i=1}^N1/k_i'$ where the times $1/k_i'$ are recursively defined by $1/k'_{N+1}=1/k_{N+1}$ and $1/k_i'=(1+k_{-i}/k'_{i+1})/k_i$ for $1\leq i\leq N$, with $1/k_i'$ representing the mean residence time in state $i$~\cite{Cao.2011}.


\end{document}